# Yoctocalorimetry: phonon counting in nanostructures

M. L. Roukes

*Condensed Matter Physics, 114-36*
*California Institute of Technology*
*Pasadena, CA 91125 USA*

It appears feasible with nanostructures to perform calorimetry at the level of individual thermal phonons. Here I outline an approach employing monocrystalline mesoscopic insulators, which can now be patterned from semiconductor heterostructures into complex geometries with full, three-dimensional relief. Successive application of these techniques also enables definition of integrated nanoscale thermal transducers; coupling these to a dc SQUID readout yields the requisite energy sensitivity and temporal resolution with minimal back action. The prospect of phonon counting opens intriguing experimental possibilities with analogies in quantum optics. These include fluctuation-based phonon spectroscopy, phonon shot noise in the energy relaxation of nanoscale systems, and quantum statistical phenomena such as phonon bunching and anticorrelated electron-phonon exchange.

# 1. Introduction

With recently developed techniques for surface nanomachining we are now able to pattern suspended, monocrystalline semiconductor nanostructures possessing extremely weak thermal coupling to their surroundings. Successive applications of these methods permit definition of quite complex devices, including those with integrated, nanoscale thermal transducers. One such process (developed by us for multilayered GaAs heterostructures) has enabled direct thermal conductance experiments upon nanostructures.[1] These advances now provide access to the domain of mesoscopic *phonon* transport — an area which, to date, has remained largely unexplored. This stands in contrast to the numerous complementary investigations in nanoscale *electronic* systems, carried out over a period spanning more than a decade.

Our new fabrication techniques permit "tuning" the energy coupling of mesoscopic structures to their environment (through geometry). This approach is beginning to enable detailed investigations of the microscopic processes through which nanometer-scale objects thermally equilibrate. In small structures at low temperatures these processes are complex, and exceedingly weak — due both to phase space restrictions arising from dimensional confinement, and to the decreased interaction strengths present in systems smaller than the length scale for bulk energy relaxation. The study of such issues is of direct relevance to nanoscale and molecular electronics. Additional motivation for this work also arises from recent theoretical efforts that establish realistic prospects for observing experimental manifestations of *thermal* conductance quantization in nanostructures.[2,3]

Here our focus is upon *calorimetry* at the nanoscale. At millikelvin temperatures the heat capacities of thermally-isolated nanostructures become extremely small. What follows is a proposal to capitalize upon this, to carry out novel explorations at minute energy scales. Crucial for such investigations is a technique for nanoscale temperature measurements that provides the requisite sensitivity while imposing minimal *back action* upon the sensitive system under study. I show that dc SQUID noise thermometry, developed in our previous work,[4,5,6] fits these criteria.

As outlined below, with this combination of approaches, exploration of heat transfer involving the exchange of individual thermal phonons appears possible. I conclude by discussing some of the intriguing experimental prospects that may emerge in the domain of quantum limited thermal transport.[7]

## 2. Description of the Nanocalorimeter.

The nanocalorimeter is conceptually depicted in Fig. 1. Its components are: *a) a phonon cavity* in the shape of a thin plate patterned from undoped *i*- (intrinsic) GaAs; *b)* four long *i*-GaAs *bridges* with rectangular cross section that support this cavity and provide very weak (phononic) thermal coupling to the environment; *c)* two *transducers* comprised of degenerate, disordered, low-density electron gases within small blocks of *n+* GaAs, patterned directly above the underlying cavity (and in expitaxial registry with it); and *d)* superconducting *leads*, running atop the bridges from the transducers to larger electrical contacts (wirebond pads) located at the supports that, in turn, are anchored to the substrate. Below $T_c$, the Nb leads provide *electrical* contact for measurements, but introduce minimal spurious *thermal* contact to the environment. Later I shall describe the measurement technique, based upon dc SQUID noise thermometry. The entire nanocalorimeter, with the exception of the Nb leads, is patterned from a single, multilayered GaAs heterostructure by a succession of electron beam lithography and pattern transfer steps. This fabrication process is described elsewhere,[1,8] topics not our main focus here.

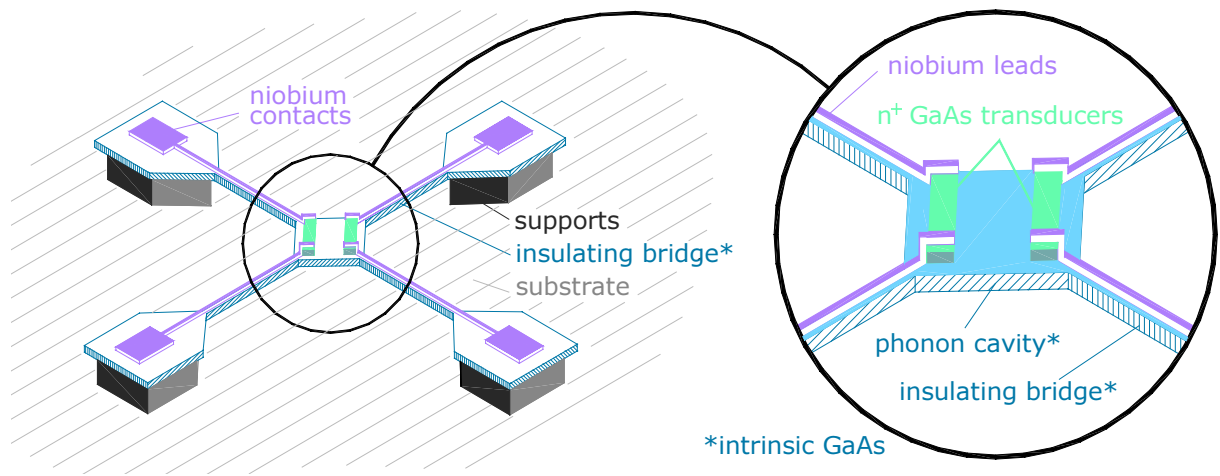

**Figure 1.** *Suspended Nanocalorimeter*. A "phonon cavity" of monocrystalline, *i* GaAs, 50-200nm thick and 1-3μm square is suspended ~1μm above the substrate (the exact standoff not critical) by four *i*-GaAs "bridges" 1-25μm long with 100nm square cross section. Degenerate, disordered electron gases within two ~100Ω, 1μm×200nm×100nm *n+* GaAs blocks, which are in epitaxial registry with the cavity, form thermal "transducers". These are electrically connected to superconducting "contacts" (wirebond pads) via ~15nm thick Nb "leads" running atop the *i*-GaAs bridges. Non-alloyed ohmic contacts between the Nb leads and the transducers are obtained by compositionally grading the latter from *n+* GaAs → *n+* InAs in their topmost ~10nm. As discussed in the text, use of two transducers facilitates both thermal transport experiments (in which one transducer is Joule heated) and time correlation measurements of thermal fluctuations (measured between the pair of unheated transducers). The dimensions given here should be considered as "representative", *i.e.* they are comparable to those of initial prototypes constructed in our laboratory. *(Ref. 8.)*

Several details are noteworthy. The bridge cross sections are narrow, *i.e.* comparable or smaller than $\lambda_{dom}$, the dominant phonon wavelength at the temperatures of the experiment.[9] Hence, at certain energies the resulting cavity geometry is, in effect, quasi-*closed*: it sustains *local* phonon modes that are only evanescently coupled to the environment through the bridges (*cf.* Fig. 8, and associated text). These bridges terminate at supports, patterned from the *i*-GaAs substrate, that are in good thermal contact with the substrate and, thereby, with a dilution refrigerator. (Together the latter two constitute the "environment".) The transducers are electrically coupled to the leads via unalloyed ohmic contacts;[10] this circumvents the need for alloyed, normal state metal contacts which would introduce a huge parasitic heat capacity (in comparison to all other components). Finally, patterning the leads from a very thin superconducting film insures that boundary scattering limited phonon conduction within them is negligible compared to that in the *i*-GaAs bridges.

### 3. Nanocalorimeter Energetics.

As is the case for bulk systems, we shall initially assume that energy exchange between the nanocalorimeter and its surroundings is reflected as a change of the calorimeter temperature, $\Delta T \sim \Delta E / C_{tot}$. We shall need to re-examine and refine the some of the assumptions implicit in this simple picture, especially for the most interesting situations involving very small occupation factors for cavity phonons. Proceeding with this initial line of reasoning, we require knowledge of three primary factors to assess the ultimate limits to the energy sensitivity of *nm*-scale calorimetry. The *first* is the temperature-dependent (total) heat capacity of the nanocalorimeter, $C_{tot}(T)$.[11] Below I shall enumerate the principal contributions to $C_{tot}$ arising from the calorimeter's independent internal degrees of freedom. The *second* is the ultimate r.m.s. temperature sensitivity, $\langle(\delta T)^2\rangle^{1/2}$, attainable for local temperature measurements within the suspended device. Ostensibly, these two factors alone determine the r.m.s. energy sensitivity of the nanocalorimeter, $\langle(\delta E)^2\rangle^{1/2} \approx C_{tot}(T) \langle(\delta T)^2\rangle^{1/2}$. However, as we shall demonstrate, a *third* factor, the back action associated with the measurement process itself, must be very carefully considered. This back action upon the calorimeter is manifested in two distinct forms: first, as a parasitic energy pathway (heat leak) from the calorimeter to the outside world (ultimately involving electrically-mediated losses within the SQUID itself);[12] and, second, as a spurious source of heating via both thermal and coherent (Josephson) radiation emanating from the measurement system. Again, since the calorimetric devices considered have minuscule heat capacity, both processes have potential to severely degrade the energy sensitivity.

I turn first to the principal contributions to the total nanocalorimeter heat capacity, $C_{tot}$. These arise from: *a)* $C_{el}(T)$, the low-density electron gases forming the two transducers, *b)* $C_{ph}(T)$, the phonons within the cavity, and *c)* from $C_{sup}(T)$, the quasiparticle excitations within the superconducting leads below $T_C$. Below, these are discussed in turn and their separate temperature dependences are plotted in Fig. 2. Later the possible existence and role of additional, spurious degrees of freedom will be considered. Such extraneous sources might arise from residual background dopants, certain types of disorder, magnetic degrees of freedom, and surface states and adsorbates.

*a)* <u>Electronic heat capacity</u>, $C_{el}$. In bulk-doped GaAs with carrier density $n \sim 2 \times 10^{18}$ cm$^{-3}$, electrons populate the bottom of the $\Gamma_6$ conduction band forming a degenerate electron gas with, to a good approximation, a spherical Fermi surface at $\varepsilon_F \sim 83$meV.[13] At this density, for the representative transducer geometry of Fig. 1, roughly $N = k_F^3 V_{tr} / \pi^3 \sim 34{,}000$ ("zero dimensional") modes are occupied. Here, $V_{tr}$ is the volume of the transducer. Impurity scattering limits the (residual) low temperature electron mobility of this material to a value $\mu$~600 cm$^2$/(V·s); this corresponds to a momentum scattering lifetime of $\tau$~$m^*\mu/e$ ~23 fs. The resulting level broadening, $\delta\varepsilon \sim \hbar/\tau$ is of order ~29 meV, and thus greatly exceeds the average energy level spacing at $\varepsilon_F$, $\langle\Delta E\rangle \sim [g(\varepsilon_F)V_{tr}]^{-1} \sim 1.5\,\mu$eV. Hence we may approximate the transducers as comprised of 3D, disordered,[14] free electron gases — *i.e.* having heat capacity that decreases linearly with temperature, $C_{el}(T) \sim 2(\pi/3)k_B^2 T g(\varepsilon_F) V_{tr}$. To a very good approximation, the density of states at $\varepsilon_F$ can be replaced by its free electron value, $g(\varepsilon_F) \sim m^* k_F/(\pi \hbar)^2$. Fig. 2 shows the temperature dependence of $C_{el}$, obtained in this manner.

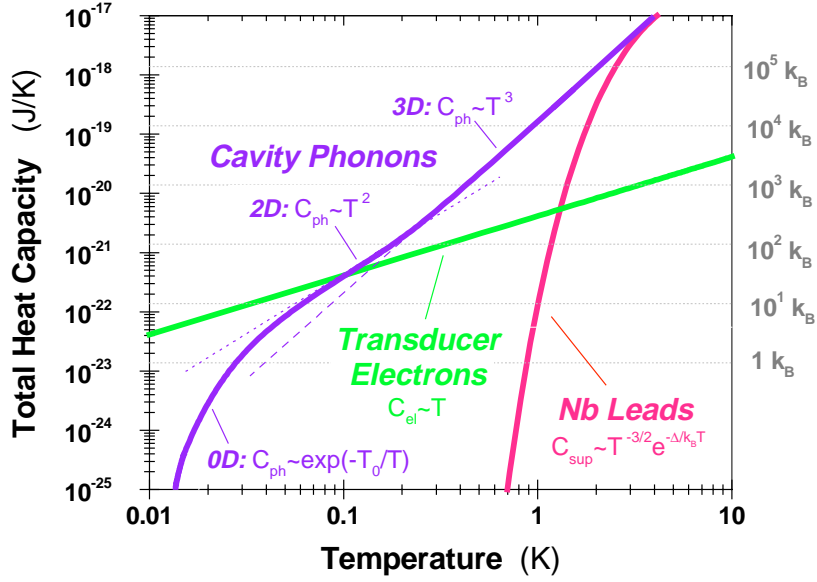

**Figure 2.** *Principal components of $C_{tot}$, the nanocalorimeter heat capacity.* For the "cavity" of Fig. 1, phonon dimensionality crosses over from 3D → 2D at ~200mK, evident here from the high temperature $T^3$ dependence (fine dashed line) giving way to a $T^2$ law (fine dotted line) at reduced temperature. At lower temperatures, here below $T$~100mK, "freeze-out" of $C_{ph}$ (from 2D → 0D) occurs. The low density electron gases forming the *n+ GaAs* transducers dominate $C_{tot}$ below ~90mK; their temperature dependence is linear. At all temperatures of interest, the heat capacity of the Nb leads is negligibly small.

*b)* <u>Phononic heat capacity</u>, $C_{ph}$. At low temperatures only the phonon modes with smallest $k$ retain finite occupation numbers. Hence we may assume phonon dispersion in the GaAs cavity is essentially linear and $c_{ph}$, its phononic heat capacity (per unit volume), is well described by the Debye model, $c_{ph}=(12\pi^4 n k_B/5)(T/\Theta_D)^3$. Here, $n$ and $\Theta_D$ are the molecular density and the Debye temperature for *GaAs* (2.2×10$^{22}$cm$^{-3}$ and 345K, respectively). However this 3D estimate, $C_{ph} \sim c_{ph} V_{cav}$ provides only an *upper* bound to the heat capacity of our system.[15] At temperatures below ~100mK, where $\lambda_{dom}$ begins to exceed all of the sample dimensions, $C_{ph}(T)$ falls exponentially. This is illustrated by the curve for $C_{ph}(T)$ displayed in Fig. 2, calculated numerically for cavity dimensions of Fig. 1 assuming linear dispersion and free boundaries.[16] Below about 100 mK, $C_{ph}(T)$ becomes exponentially suppressed, *i.e. zero* dimensional, since the

**6**

dominant, smallest $q$ modes behave as Einstein oscillators below their characteristic temperature, $T < \hbar v_s q / k_B$.

*c)* <u>Heat capacity of superconducting leads.</u>  As depicted in Fig. 9, the quasiparticle contribution to the specific heat of a BCS superconductor is exponentially suppressed below $T_c$.[17]  For $T<<T_c$,

$$C_{sup}(T) = 3Nk_B \left(\frac{\pi}{2}\right)^{1/2} \frac{\Delta}{\varepsilon_F} \left(\frac{\Delta}{k_B T}\right)^{3/2} \exp\left(\frac{-\Delta}{k_B T}\right)$$

Only above ~2K will these quasiparticle excitations rise sufficiently to affect $C_{tot}$.

Figure 2 shows that, below ~90mK, the temperature dependence of $C_{tot}$ becomes dominated by the contribution from the electron gases comprising the transducers.  For the prototypical geometry of Fig. 1 the total calorimeter heat capacity in this regime is then $C_{tot} \sim (2.2 \times 10^{-21}$ J/K$^2) \cdot T$ i.e. ~ $(160 \, k_B/K) \cdot T$. This yields $C_{tot} < 2 \, k_B$ at a base temperature of $T=10$mK.

I now evaluate $<(\delta T)^2>^{1/2}$, the temperature resolution available from dc SQUID noise thermometry. This can be expressed as $<(\delta T)^2>^{1/2} \sim \sigma \, (T + T_{sys})$, where $T$ and $T_{sys}$ are the ambient (transducer) and "effective system noise" temperatures, respectively, whereas $\sigma$ is the fractional measurement precision.  The latter depends upon both $\Delta\omega$, the measurement bandwidth, and $t_{meas}$, the measurement integration time, viz. $\sigma^2 = 2/(\Delta\omega \, t_{meas})$.[18]  For the transformer coupled system of Fig. 6, $T_{sys} = \varepsilon_c \, \kappa_t \Delta\omega \, / k_B$;[4] here $\varepsilon_c$ is the coupled energy sensitivity of the SQUID and $\kappa_t$ is the degradation factor characterizing the input coupling network (typically ~5 in our systems). Previously,[5,6] using very low noise thin film dc SQUIDS cooled to 4.2K ($\varepsilon_c$~63$\hbar$),[19] we demonstrated $T_{sys}$ ~100μK. This enabled a temperature resolution $<(\delta T)^2>^{1/2}$ ~100μK at $T=10$mK with $t_{meas}$~0.3 s.  Ketchen has since reported devices yielding $\varepsilon_c$ ~3$\hbar$ when cooled to 300mK.[20]  Such advanced SQUIDs permit high sensitivity measurements, $<(\delta T)^2>^{1/2}$ ~1mK at $T=10$mK ($\sigma$ ~0.1), with much wider bandwidth, $\Delta\omega/2\pi$ ~1.4MHz, and, hence, yield short averaging times, $t_{meas}$~20 μs. This will prove essential for the present work.

Combining these two factors now allows us to arrive at an initial estimate for the energy resolution of the nanocalorimeter, valid for the case where back action is insignificant. (We examine this issue in detail below.) At $T=10$mK, a temperature resolution $\langle(\delta T)^2\rangle^{1/2} \sim 1$ mK yields an energy resolution of $\langle(\delta E)^2\rangle^{1/2} \sim C_{tot}(T) \langle(\delta T)^2\rangle^{1/2} \sim 2.2 \times 10^{-26}$ J, *i.e.* $\sim 1.4 \times 10^{-7}$ eV (140 neV).[21] This rises only to 2.8 µeV at $T=100$mK, a value that includes the additional contribution from the phonons at this temperature.

## 4. Nanocalorimeter Dynamics: Phonon Counting.

At 10mK the average energy of a thermal phonon is $\hbar\omega \sim k_B T \sim 0.9$ µeV. In the previous discussion I have shown that nanocalorimetry is feasible with energy sensitivity $\langle(\delta E)^2\rangle^{1/2} \sim$ 140neV. Detection of individual phonon exchanges thus appears possible, at least in terms of the *energetics* involved; but the analysis so far says nothing about *temporal* considerations. To explore this issue we shall now turn to the phonon arrival rate, and to the cascade of energy exchange processes involved in "phonon counting".

A conceptual diagram of energy flow within the nanocalorimeter is presented in Fig. 3. Consider the "deposition" of a phonon from the environment into the nanocalorimeter. The transmission probability into the cavity is determined by the bridges and by the details of their coupling at both ends.[22] "Counting" a phonon upon its arrival into the cavity requires its annihilation, via the electron-phonon interaction between the cavity and transducer. This process creates a hot electron in the transducer. Electron-electron scattering then acts within the transducer to re-thermalize the carriers, yielding a new, slightly hotter, electron temperature.

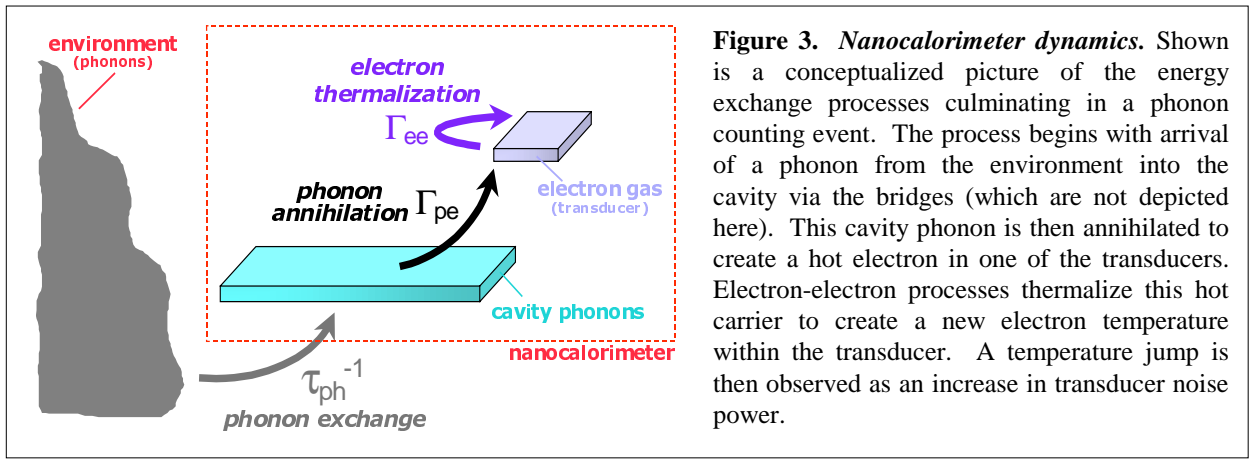

**Figure 3.** *Nanocalorimeter dynamics.* Shown is a conceptualized picture of the energy exchange processes culminating in a phonon counting event. The process begins with arrival of a phonon from the environment into the cavity via the bridges (which are not depicted here). This cavity phonon is then annihilated to create a hot electron in one of the transducers. Electron-electron processes thermalize this hot carrier to create a new electron temperature within the transducer. A temperature jump is then observed as an increase in transducer noise power.

This increases the magnitude of the transducer's thermally-driven electron density fluctuations, with result that a slightly larger Nyquist noise level is read out by the dc SQUID.[23] For this sequence of events to culminate in phonon counting, the rates must be in the correct proportion: the phonon arrival rate, $\tau_{ph}^{-1}$, must be small compared to $\Gamma_{pe}$, their rate of annihilation within the cavity, and the latter, in turn, must be slow compared to the thermalization rate for hot electrons within the transducer, $\Gamma_{ee}$. Finally, the measurement integration time, $t_{\text{meas}}$ must be on the scale of $\tau_{ph}^{-1}$, or shorter. If this hierarchy holds, every phonon that arrives (or that leaves, by reciprocity) results in a discrete jump of the electron *temperature* measured at the transducer. I now turn to a preliminary assessment of these rates.

*a)* <u>Phonon exchange rate</u>, $\tau_{ph}^{-1}$. As mentioned, coupling between cavity phonons and those in the environment is mediated by the bridges. At low temperatures, $T \ll \Theta_D$, thermal transport becomes limited by boundary scattering. For a nanocalorimeter in which phonon surface scattering is *specular*, estimation of $\tau_{ph}^{-1}$ is rather complicated. In this case it depends on the details of mode coupling between the calorimeter and the environment — *i.e.*, upon the detailed mode structure of the nanocalorimeter (coupled modes involving bridges + cavity), and also upon the precise geometry of the region connecting the bridges to the supports. I shall take a different, simpler approach here to obtain an initial estimate of $\tau_{ph}^{-1}$ by assuming that phonon surface scattering within the bridges is *diffuse*. (This is consistent with results from our initial nanoscale thermal conductance measurements.)[24] For this case, the thermal conductance of the four bridges is given by $G_{ph} \sim (4/3) c_{ph} v_{av} \Lambda (A_{\text{bridge}}/L_{\text{bridge}})$, where $L_{\text{bridge}}$ and $A_{\text{bridge}}$ are the bridge length and cross sectional area, $v_{av}$ is the polarization-averaged phonon velocity, and $c_{ph}$ is the (3D) Debye heat capacity per unit volume; *i.e.* $C_{ph} = c_{ph} V_{\text{cav}}$, where $V_{\text{cav}}$ is the cavity volume. Consistent with our assumption, the phonon mean free path within the bridges, $\Lambda$, is assumed temperature independent and of order $\sim \sqrt{A_{\text{bridge}}}$.[25] The phonon exchange rate[26] is then also temperature independent, and of order $\tau_{ph}^{-1} \sim G_{ph}/c_{ph} V_{\text{cav}} \sim 4 v_{av} A_{\text{bridge}}^{3/2} / (3 L_{br} V_{\text{cav}})$. To give a concrete example (*cf.* Fig. 1), consider a 3μm×3μm×50nm *i*-GaAs cavity, connected by 25μm×50nm×50nm bridges to the environment. This example, which is at the limit of our current technology (given the extreme aspect ratio of the bridges, ~500), yields $\tau_{ph}^{-1} \sim 5\times10^4$ s$^{-1}$.

*b)* <u>Phonon annihilation rate</u>, $\Gamma_{pe}$. For small energy transfer, *i.e.* small disequilibrium between the phonons and the transducer electrons, we may estimate the phonon-electron energy transfer rate as being essentially equal to the phonon-electron scattering rate in thermal equilibrium. Ziman[27] has calculated the mean free path for Debye phonons interacting with a free electron gas (parabolic band) at low temperatures — within a simple deformation potential model relevant to our system. Again I neglect, in this first approximation, scattering-phase-space reduction at low temperatures (we shall return to this issue below). For bulk phonons and electrons coexisting in thermal equilibrium within *n+ GaAs* at *n*~2×10$^{18}$ cm$^{-3}$, employing the *GaAs* deformation potential constant of 7eV,[28] Ziman's model implies $\Gamma_{pe} \sim (2.14\times10^7$ s$^{-1}\cdot$K$^{-1})\cdot T$. At 1K this indicates that electrons respond in about 50ns to a (hypothetical) abrupt change in phonon temperature; at 100mK this response time increases only to about 0.5μs.

*c)* <u>Electron thermalization rate</u>, $\Gamma_{ee}$. Again for the case of weak disequilibrium, here between a hot electron and the (thermally-distributed) transducer electrons, we may approximate the electron thermalization rate as being essentially equal to the electron-electron scattering rate in thermal equilibrium. At low temperatures, the transducers— being disordered, nanoscale electronic systems — exhibit either 2D or 3D corrections to bulk rates. The exact dimensionality depends upon the ratio of the phase breaking length, $\ell_\varphi$, to sample dimensions. Measurements in *n+* GaAs indicate that $\ell_\varphi$ becomes comparable to the transducer dimensions (Fig. 1) below *T*~1K.[29] Hence we consider rates for both 2D and 3D systems: using the transducer parameters yields, in the former case[30] $\Gamma_{ee}^{(2D)} \sim (3.4\times10^{10})\ T\ \ln(3.6\times10^7/T)\ [$s$^{-1}]$; whereas, in the latter case[31] $\Gamma_{ee}^{(3D)} \sim (5.4\times10^8)\ T^2 + (8.4\times10^8)\ T^{3/2}\ [$s$^{-1}]$. (In these expressions T is to be expressed in Kelvin.) For both dimensionalities these electron-electron scattering rates greatly exceed the rates for phonon exchange and annihilation over the entire temperature range of relevance (*cf.* Fig. 5).

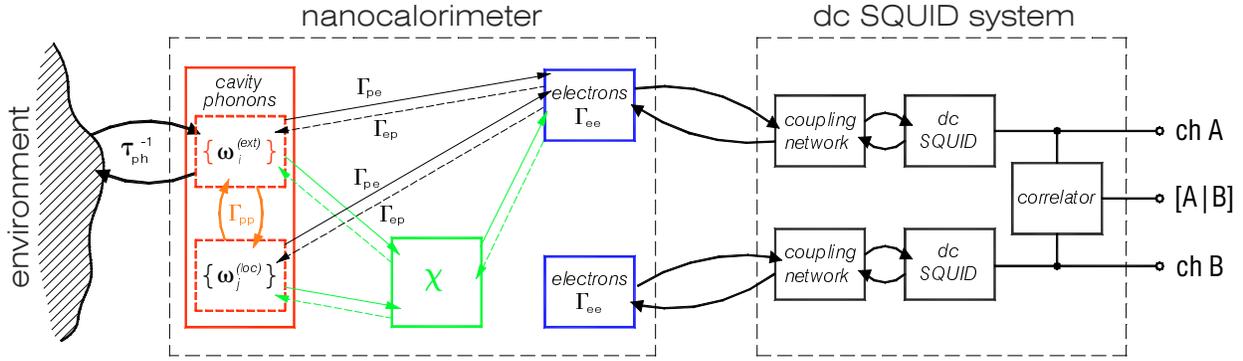

**Figure 4.** *Nanocalorimeter degrees of freedom and energy pathways*. The large dashed boxes delineate the constituents of the nanocalorimeter *(left)* and the two-channel dc SQUID measurement system *(right)*. The smaller boxes at the left represent the nanocalorimeter's individual DOF, with arrows depicting the scattering mechanisms that couple them. In the case of specular surfaces, phonons are differentiated into classes of extended, $\{\omega_i^{(ext)}\}$, and localized modes, $\{\omega_j^{(loc)}\}$, as described in the text and Figure 8; only the former have direct, efficient coupling to the environment. (The internal couplings are displayed for one transducer ("electrons"); the other's are identical.) The box labeled $\chi$ represents either spurious DOF, or intentional surface adsorbates (see text). The smaller boxes at the right represent internal components of the measurement system: two dc SQUIDs coupled to the transducers via frequency-dependent networks. A correlator enables temporal cross-correlations between the two output "channels".

These initial estimates illustrate that the rates are, indeed, correctly ordered to permit phonon counting. We turn now to a more detailed consideration of energy exchange processes within the nanocalorimeter.

## 5. Scattering Rates: Further Discussion.

The principal energy reservoirs involved in nanocalorimetry are depicted in Fig. 4. Rough estimates for the scattering processes coupling these degrees of freedom are provided in Fig. 5. Several points are noteworthy. The rates plotted have been obtained by applying *bulk* formulae (scaled to the nanocalorimeter's parameters, *e.g.* geometry, electron density). Effects of reduced dimensionality, in both the phonon density of states and in the coupling between the electrons and phonons, are not included — more accurate rates can be obtained only through detailed calculations, explicitly tailored to these nanoscale systems. Accordingly, Fig. 5 should be viewed as providing a qualitative initial guide to the relevant microscopic rates and processes. To date, very little theoretical work has been carried out in this area, it is clearly crucial to these issues.

In Fig. 5, Trace 1 represents the temperature dependence of the *system* energy equilibration rate, $\tau_{tot}^{-1} = C_{tot}/G_{ph}$, which tracks the slowest rate amongst the series of processes leading to system equilibration. Trace 2 shows the phonon exchange rate, $\tau_{ph}^{-1}$, as defined previously. Trace 3 depicts $\Gamma_{ep}$, the acoustic phonon emission rate from a hot, three-dimensional electron gas — assuming electrons and phonons are Fermi and Bose distributed, respectively.[32] The $T^3$ power law displayed is obtained arises from bulk deformation potential theory. Trace 4, and Traces 5a and 5b depicts the phonon annihilation rate, $\Gamma_{pe}$, and both the 2D and 3D electron-electron scattering rates in the presence of disorder, each obtained as discussed previously. Not shown is the phonon-phonon scattering rate arising from lattice anharmonicity; for $T<<\Theta_D$ this is negligibly small compared to all other processes.[33]

Several additional points are worthy of mention. At the lowest temperatures, the difference between $\tau_{tot}^{-1}$, the rate at which the entire nanocalorimeter equilibrates, and $\tau_{ph}^{-1}$, the phonon exchange rate, reflects the huge difference between the electronic and phononic heat capacities. In this regime, in equilibrium, thermal fluctuations of the two transducers are uncorrelated. The situation changes at the upper end of the temperature range, where $\tau_{ph}^{-1}$ becomes small compared to both $\Gamma_{ep}$ and $\Gamma_{pe}$. In this regime the transducers become more strongly coupled to each other (via their interaction with cavity phonons) than they are to the environment. We shall return to this point shortly. Note that to register an individual annihilation event the transducer must remain at elevated temperature over the measurement time; hence $\Gamma_{ep}^{-1} \geq t_{meas}$. In the bulk, this holds below $T \sim 150$ mK for state-of-the-art SQUIDS ($t_{meas} \sim 20\mu s$).

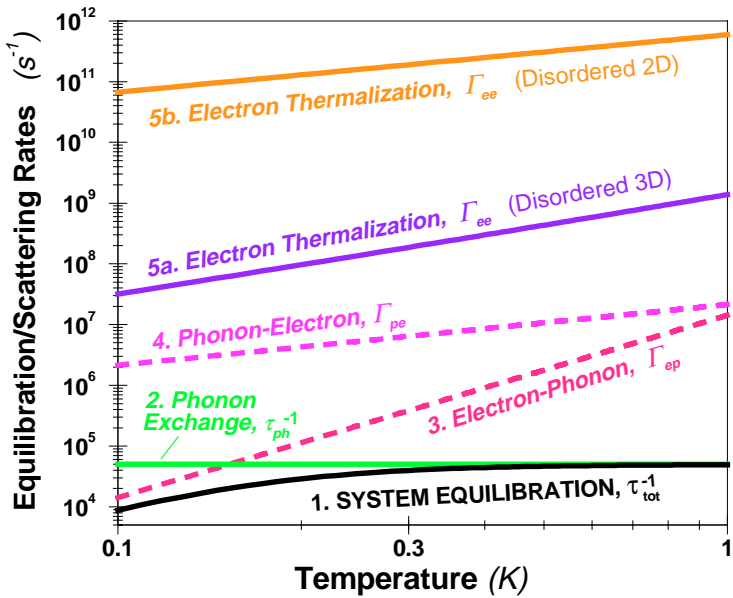

**Figure 5.** *Estimated equilibration and scattering rates for the nanocalorimeter.* Between 100mK and 1K, the transducers respond, and internally thermalize, more quickly than the system exchanges energy with its environment. For *T*>220mK temporal correlations between the sensors are expected; in this regime, phonon-mediated energy exchange between the transducers occurs at a faster rate than phonons are exchanged with the environment. Effects arising from reduced phonon dimensionality and size dependence in the electron-phonon coupling have not been included in these simple estimates; these are expected to become important for T<100mK.

## 6. SQUID Back Action upon the Nanocalorimeter.

If unaddressed, back action from the dc SQUID measurement system can potentially compromise these very low energy measurements in two distinct ways. First, since a dc SQUID is operated in the finite voltage state, a real component of its input impedance exists. This can dissipate energy stored within the calorimeter. Second, both high frequency Josephson radiation and directly-coupled, low frequency circulating current noise of the SQUID can affect the transducers. Their r.m.s. levels can result in dc heating, while the low-frequency a.c. components can directly drive the electron gas within the measurement band. The magnitude of these terms depends upon the frequency-dependent coupling between the calorimeter and the SQUID, as well as the SQUID's dynamic characteristics.[34] These effects can be minimized with careful design.

*Parasitic energy loss.* The transformer-coupled input circuit (Fig. 6) is completely superconducting except for the *n*+ GaAs transducer, and the dissipation intrinsic to the SQUID loop itself, which I shall denote as $\mathcal{R}$. Theoretical analyses of $\mathcal{R}$[35,36,37] and one measurement in a modern, high sensitivity thin-film dc SQUID[38] indicate that $\mathcal{R} \sim R_{sh}$ at optimal flux and current bias conditions. Here $R_{sh}$ is the resistance of normal shunts placed across each Josephson junction to obtain non-hysteretic response. To estimate the dissipation they introduce (as viewed from the perspective of transducer), first consider the hypothetical case where a resistor, $R_1$, held at temperature $T$, is coupled to a second resistor, $R_2$, held at temperature $T=0$. In the limit $R_2 \gg R_1$, for coupling mediated by a frequency-dependent network with roughly unity transmission over bandwidth $\Delta f$, the power transferred from $R_1$ to $R_2$ is $P_{1\rightarrow 2} \sim 4k_BT\,\Delta f\,(R_1/R_2)$. Applying this to the case at hand, I equate $R_1$ with the transducer ($\sim$100Ω) whereas $R_2 = [\omega \mathcal{M}\, M/L_{in}]^2/\mathcal{R}$ is the effective SQUID dissipation at $T=0$, reflected back through the input circuitry. Here $\mathcal{M}$, M and $L_{in}$ are defined in Fig. 6. Typical values for our dc SQUIDS and transformers are: $\mathcal{R} \sim R_{sh} \sim 2\Omega$, $\mathcal{M} \sim 7\mu H$, M$\sim$10nH, and $L_{in} \sim 1\mu H$. Integrating the resulting expression for $P_{1\rightarrow 2}(\omega)$ up to the (typical) self-resonance of the transformer ($\omega_{res} \sim 10^6\,s^{-1}$), yields $P_{1\rightarrow 2}\cdot t_{meas}/4k_BT \sim 3\times 10^{-4}$, indicating that energy transferred over the measurement time of the coupled calorimeter is roughly four orders of magnitude smaller than the thermal energy.

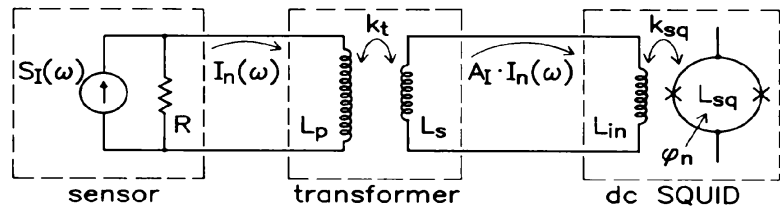

**Figure 6.** *Transformer-coupled SQUID system.* The mutual inductances for the superconducting transformer and dc SQUID are $\mathcal{M} \sim \sqrt{(k_t^2 L_p L_s)}$ and $M \sim \sqrt{(k_{sq}^2 L_{in} L_{sq})}$, respectively. The shunt resistors across each junction (denoted by ×) are not displayed. After *Roukes et al.*, Ref. 4.

*Thermal and Josephson Radiation*. The circulating current noise (spectral density $S_J$) within the SQUID loop can couple power back into the electron gas of the sensor. The magnitude of $S_J$ is known for modern dc SQUIDs through simulations[39] and direct measurement[40]. In general, $S_J = \gamma_J(4k_BT/2R_{sh})$, where $\gamma_J$ is the factor by which the circulating current noise exceeds (equilibrium) Nyquist noise expected from the shunt resistors. Theory and experiment yield $\gamma_J \sim$ 6-11 at optimal bias conditions. To estimate this form of back action, we assume $\gamma_J \sim 10$ and $T_{squid}$=100mK (electron heating typically occurs in dc SQUIDS below this temperature).[41] Referring $S_J$ back through the input circuitry yields, *at the transducer*, $S_I^{(eff)} \sim \int d\omega\,\mathcal{F}(\omega)\,S_J$, where the reverse coupling is characterized by the factor $\mathcal{F}(\omega) = \omega^2[(\mathcal{M}\,M)/(L_{in}R)]^2/[1+(\omega\tau_p)^2]$. Here $\tau_p = L_p/R$; all other symbols are defined in Fig. 6. The integral is again taken up to $\omega_{res}$. For $L_{in}\sim$ 1μH, we obtain $\mathcal{F}(\omega) \sim [1.2\times 10^{-17}\,sec^{-2}]\cdot\omega^2/[1+(\omega\tau_p)^2]$. This yields an effective back action noise term, $I_{rms} \sim 1.3\times 10^{-12}$ A, hence the power delivered to the transducer over $t_{meas}$ is only a small fraction of the thermal energy, of order $I_{rms}^2 R_{sensor}\,t_{meas}/k_BT \sim 1.2\times 10^{-4}$ for $T\sim$10mK and $T_{squid} \sim$100mK. This fraction only rises to $\sim 5.1\times 10^{-3}$ for $T_{squid}$=4.2K.[42]

Finally, the dc SQUID operated in the finite voltage state produces Josephson radiation at frequency $f = eV_b/\pi\hbar = (484\,MHz/\mu V)\,V_b$, where $V_b$ is the SQUID bias voltage. For our SQUIDs, $V_b$ is in the range of 10 – 100 μV, hence suppression of subnanowatt level coherent radiation occurring in a band from 4 – 40 GHz is necessary. These are much higher frequencies than the band used for fluctuation measurements, hence placing frequency-dependent networks between the transducer and the dc SQUID can suppress this microwave radiation, while still allowing the requisite coupling to the transducer's low frequency fluctuations.[43]

## 7. "Spurious" Heat Capacity.

The projected low temperature heat capacity of the calorimeter approaches a few $k_B$ at 10mK (Fig. 2). Clearly additional, parasitic degrees of freedom (DOF) could significantly increase this value. These could include two-level systems (TLS) arising from disorder in the bulk or on the semiconductor surfaces, nuclear Zeeman and quadrupole DOF, and DOF arising from surface adsorbates. However, the mere presence of such additional DOF does not necessarily imply that the sensitivity of the calorimeter will be compromised. Two factors are relevant: the *activation energy* required to access a particular DOF, and the *rate of energy exchange* between DOF directly employed in calorimetry ($C_{el}$ and $C_{ph}$), and those that are parasitic. Figure 7 illustrates the first point. The contribution to parasitic DOF involving a single TLS with an energy level separation, $\Delta E$, falls off rapidly both well above and well below the activation temperature, $T_a = \Delta E/k_B$. For $T \ll T_a$, the TLS heat capacity is exponentially suppressed, $C(T)/k_B \sim (\Delta E/k_B T)^2 \exp(-\Delta E/k_B T)$. For $T \gg T_a$, the decay with temperature is weaker, $C(T)/k_B \sim (\Delta E/2k_B T)^2$. Processes with activation energies within the temperature regime of interest have the greatest potential to increase $C_{tot}$. But the second point is that even these DOF are harmless if energy exchange between them and the principal calorimeter DOF occur on a much slower time scale than that of the experiment itself, $t_{meas}$. In this situation, the spurious degrees of freedom simply contribute a slowly drifting background (a "time-dependent heat leak") to the measurements. If very slow compared to the measurement time, their effect can be separated from the desired data.

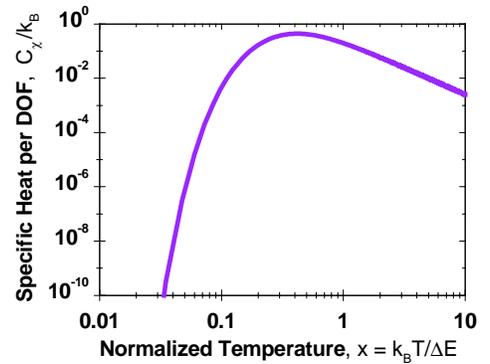

**Figure 7.** *Heat capacity for a single TLS DOF.* $C_\chi(T)$ quickly decays away from its maximum at $T \sim 0.39 \Delta E/k_B$ — exponentially at the low $T$ end, and as $x^{-2}$ at the high end.

Excess heat capacity, beyond that expected for crystalline systems, has been linked to TLS in disordered glasses.[44] Millikelvin temperature measurements of $Q^{-1}$ in *Si* mechanical oscillators have identified shallow acceptor states — randomly split in energy due to random local strain fields — as one microscopic origin of such low energy TLS in high purity material.[45] No similar body of work appears to exist for *GaAs*; hence, similar estimation of the density of TLS in our devices is difficult. However, our suspended samples are fabricated from ultrapure *GaAs*, typically with unintentional *p*-type background doping $\sim 1\times 10^{14}$ cm$^3$. This decreases to *below* $\sim 2\times 10^{13}$ cm$^3$ in the best heterojunctions available to us.[46] For $V_{tot}= 4\times 10^{-13}$ cm$^3$, literally just a few impurity centers are present — these can only affect $C_{tot}$ below the temperature regime of the experiments, $T<10$mK.[47]

We turn now to the nuclear DOF. The energy level spitting for a given nuclear species, *i*, in an ambient field $H$ is $\Delta E_i = \hbar \gamma_i H$; here $\gamma_i$ is the gyromagnetic ratio for species *i*. For the nuclei of atoms comprising the calorimeter, $\gamma_i$ range from $4.58\times 10^3$ (As$^{75}$) to $8.16\times 10^3$ G$^{-1}\cdot$s$^{-1}$ (Ga$^{71}$). These yield activation temperatures $T_a = \Delta E_i/k_B \sim 10^{-8}$ K in the earth's field ($\sim 0.3$G). For experiments at $T>10$mK the total nuclear heat capacity of the calorimeter is $C_{nuc}(T) \sim k_B \cdot (H/T)^2 \cdot \Sigma_i \{ \eta_i \rho_i V_i (\hbar \gamma_i / 2k_B)^2 \}$. Here $\eta_i$, $\rho_i$, and $V_i$ represent, respectively, the natural abundance and the (isotopic average) mass density of species *i*, and the volume of the specific part of the calorimeter containing species *i*.[48] Summing over contributions from the nuclear species present yields $C_{nuc}(T) \sim k_B (H/T)^2 (1.1\times 10^{-5}$ G$^{-2}$K$^2)$. At $\sim 0.3$G and $T\sim 10$mK this yields $C_{nuc}(T) \sim 0.01 k_B$.

In principle, unpaired electron spins could pose a more serious problem given that their magnetic moments are larger by a factor of $10^3$. For a worst case scenario, we assume that shallow acceptors, at a background density of $1\times 10^{14}$ cm$^{-3}$ (our upper bound) within the *i* GaAs, are completely *un*ionized (*i.e.* neutral) at $T\sim 10$mK. For $g=2$, $\Delta E_e = \hbar \gamma_e H$, where the free electron gyromagnetic ratio, $\gamma_e$, is $1.76\times 10^5$ G$^{-1}\cdot$s$^{-1}$. Hence, in the earth's magnetic field we obtain $T_a=0.4\mu$K. The low density and the small cavity volume imply that, at most, only $\sim 50$ of such species are present, these yield $C_{\text{el-spins}}(T) \sim k_B \cdot (H/T)^2 \cdot (2.1 \times 10^{-11} \text{ G}^{-2}\text{K}^2)$, *i.e.* only about $\sim 2 \times 10^{-8} k_B$ in the earth's field at 10mK.

Surface DOF could be present in two principal forms: shallow electronic traps near the depletion regions at the edges of the *n+ GaAs* electron gases, and two-level systems at all surfaces due to unknown electronic states or surface adsorbates. Below, I address each in turn.

The *n+ GaAs* transducers in our prototype calorimeter have surface area of $5.8 \times 10^{-9}$ cm$^2$. Their doping of $n \sim 10^{18}$ cm$^{-3}$ implies a "surface density" of about $n_s \sim 10^{12}$ cm$^{-2}$ distributed within a surface layer of thickness about 10nm. There will actually exist a surface depletion layer of depth ~30nm given the a typical surface state density ~$10^{10}$ cm$^{-3}$. Let us assume that the electrochemical potential recovers to its bulk value over a region of order 10nm. Then ~6000 donors are on the threshold of recapturing carriers in this region. Let us further assume that these constitute TLS spread uniformly distributed in energy over ~1eV (*i.e.* up to the ionization energy). Only those processes with energies in the range 10mK-1K (*i.e.* 0.9μeV-90μeV, an energy range of order 10$^{-4}$ eV) are of concern. Only of order *one* such TLS, will be situated energetically to contribute to $C_{\text{tot}}$. I anticipate that the electronic states literally at the transducer's surfaces will have $\Delta E$ ~1eV, and should not contribute to $C_{\text{tot}}$ at low temperatures.

It is difficult to obtain definitive *a priori* estimates for all possible "parasitic" degrees of freedom that might thwart attainment of the ultimate sensitivities. However, the high resolution heat capacity measurements required to optimize nanocalorimeter performance will lead to interesting research pursuits in their own right. I illustrate several possible examples below.

## 8. Nanocalorimetry and Surface Adsorbates.

Studies of the adsorption of *individual* atoms or molecules on surfaces is feasible with a nanocalorimeter at low temperatures. Most adsorption processes have binding energies, $E_B$, that greatly exceed the calorimeter's projected noise floor. For example, *He* on many bare surfaces has a typical binding energy of order $E_B/k_B$ ~10-100K.[49] One of the lower values of $E_B$ is that of

$^4$He upon a $^4$He layer, with $E_B/k_B \sim 7$K ($E_B \sim 600\mu$eV)[50]. However, even this is more than three orders of magnitude above our projected noise floor at 10mK of ~140 neV. It seems clear that, well before the requisite sensitivity to permit phonon counting is achieved, novel heat capacity measurements will become possible. Two experimental possibilities, involving thermally-activated surface diffusion and the energy relaxation following surface adsorption, are outlined below.

On perfectly smooth surfaces, adsorbed gases with submonolayer coverage should form energy bands within which, in the dilute limit, the atoms move as if an ideal 2D gas.[51] Disorder sullies this pristine picture through its introduction of random pinning potentials for the adsorbates — with result that the lateral motion of the adsorbates becomes *diffusive*. Nanocalorimetry may offer a unique means of observing the kinetics of such surface diffusion processes. For example, stochastic bursts of heat emitted when diffusing adsorbates undergo phonon-assisted mechanical rearrangement process may be measurable. Alternatively, it may be possible to study the energy spectrum of the stochastic pinning potential by comparing the heat capacity of calorimeters (plus adsorbates) having surfaces that are smooth against those intentionally roughened in a quantified manner.

Physical rearrangement of adsorbates on a rough surface constitutes an energy reservoir of two-level systems. Let us ask what coverage of such two-level systems is necessary to add ~ $1k_B$ to $C_{tot}$. (This consideration is of relevance concerning "spurious" DOF.) Approximating these atoms as having activation energies distributed uniformly from zero up to $E_B$ (assume $^4$He with $E_B/k_B \sim 100$K), leads one to the conclusion that only a fraction $\sim 10^{-2}$ contribute to $C_{tot}$ over the ~1K energy range of importance. Since a full monolayer corresponds to surface density $n_s \sim 3 \times 10^{14}$ cm$^{-2}$, complete coverage of the calorimeter (surface area $\sim 9 \times 10^{-8}$ cm$^2$) involves $\sim 3 \times 10^7$ atoms. Therefore only $\sim 3 \times 10^5$ are active in the sense of being laterally pinned within the "right" range of values of $E_B$. This crude estimate suggests that coverages below $10^{-4}$ monolayers will be required to avoid compromising the ultimate calorimeter's sensitivity. This should be possible dynamically, via techniques for desorption of surface species *in situ* at low temperatures — by making use of the intrinsically ultrahigh vacuum environment within a mK cryostat in conjunction with locally applied heat to the nanocalorimeter's surface.[52]

High resolution nanocalorimetry may also enable observation surface adsorption *dynamics*. The bandwidth of SQUID systems is insufficient for observing the initial vibrations of physisorbed species ($\omega \sim E_B/\hbar \gg 1$GHz). However, µs-scale temporal measurements, which are possible, may give insights into the thermalization pathways by which an arriving species transfers its vibrational energy to the surface after adsorption. These are likely to become quite slow at mK temperatures where reduced phonon dimensionality impairs the coupling between adsorbates and the surface.

### 9. Phonon Transport at the Quantum Limit: Equilibrium Fluctuations.

At temperatures where phonon arrival occurs on a time scale longer than the response time of the nanocalorimeter, phonon counting becomes possible. This is a consequence of the fact that the energy of a single thermal (cavity) phonon, $\hbar\omega_i$, is sufficient to raise the (transducer) electron temperature by an observable amount in this regime, $(\Delta T)_i = \hbar\omega_i / C_{el}(T) \geq \langle(\delta T)^2\rangle^{1/2}$. Here, *i* represents the mode index for cavity phonons. "Observation" of a cavity phonon is thus directly linked to its annihilation. However, it is only at the lowest temperatures, when the thermal occupation numbers for cavity phonon modes, $\langle n_i \rangle = [\exp(\hbar\omega_i / k_B T) - 1]^{-1}$, become exponentially small,[53] that a one-to-one correspondence between an upward (or downward) jump in transducer temperature and the absorption (or transmission) of a cavity phonon from (or to) the environment is most likely. At elevated temperatures, when the $\langle n_i \rangle$ exceed unity, phonon annihilation/emission events of the transducer electron gas are not, in general, correlated with phonon exchange between the cavity and the environment.[54] In the low temperature regime, these jumps in electron temperature enable a direct spectroscopy of the cavity modes. Prior to discussing this, however, we first need to explore the nature of the cavity phonon spectrum.

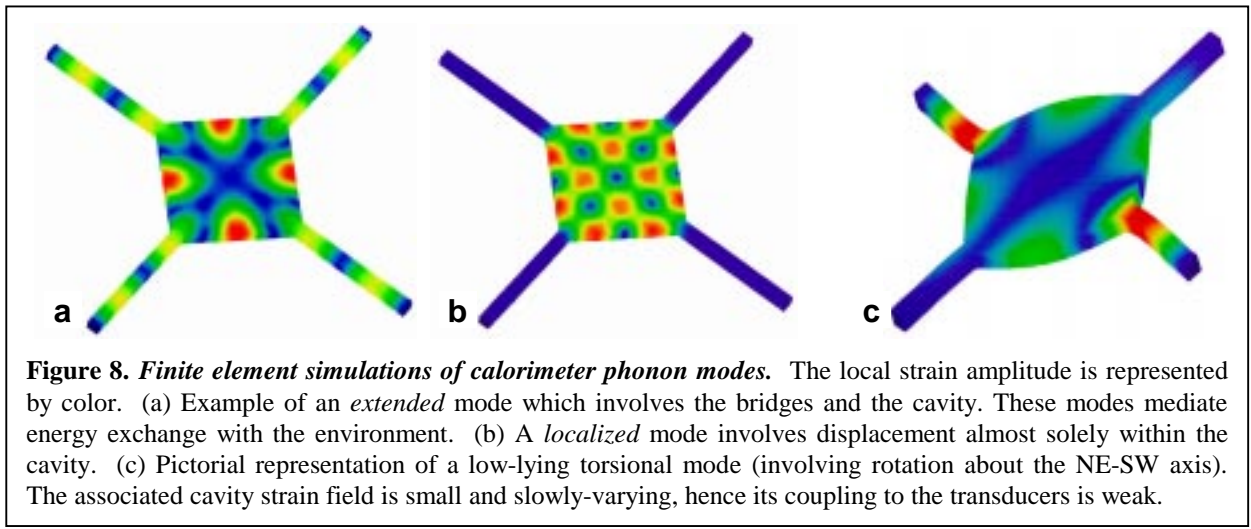

**Figure 8.** *Finite element simulations of calorimeter phonon modes.* The local strain amplitude is represented by color. (a) Example of an *extended* mode which involves the bridges and the cavity. These modes mediate energy exchange with the environment. (b) A *localized* mode involves displacement almost solely within the cavity. (c) Pictorial representation of a low-lying torsional mode (involving rotation about the NE-SW axis). The associated cavity strain field is small and slowly-varying, hence its coupling to the transducers is weak.

When phonon boundary scattering is *specular*, the phonon modes of the nanocalorimeter — considered as a single coupled system (cavity + bridges) — can be differentiated into *extended* and *localized* (Fig. 8). Only the former are well-coupled to the environment; the latter have appreciable lattice displacement solely within the cavity, while their amplitude decays in the bridges. A class of low energy torsional and "trampoline" modes of the calorimeter also exists, as exemplified in Fig 8c.[55] These are expected to be very poorly coupled to the transducers, due to extremely unfavorable overlap between spatial factors (mode shapes). Thus, at low temperatures, $T \ll \Theta_D$, where phonon-phonon scattering becomes negligible, the existence and relative occupancy of such modes should have little bearing upon the experiments.[56] Note however that, for $T \ll \Theta_D$, despite the absence of direct phonon-phonon coupling via lattice anharmonicity, such inelastic coupling *is* mediated by transducer electrons. In general, this holds for both extended and localized phonon modes. This simple picture for the cavity mode spectrum changes when phonon boundary scattering within the nanocalorimeter is *diffuse*; in this situation the actual phonon mode shapes do not possess the well-defined relation to the cavity geometry as is exemplified in Fig. 8a and 8b.

If discrete jumps in the sensor temperature are resolvable, they will provide a novel means of energy spectroscopy. The temperature jump amplitudes, $(\Delta T)_i$, each correspond directly to the energies, $\hbar\omega_i$, of active cavity phonon modes. *All* phonon modes with coupling to the transducers participate, both extended and localized. The likelihood of observing a given mode depends, for the case of absorption, upon its respective occupation factor and, for emission, upon the degree of disequilibrium existing between the transducer electrons and cavity phonons. Thus, the $\langle n_i \rangle$ can be deduced by constructing histograms of event probability vs. specific jump size. However, such interpretation may require knowledge of the (modal) electron-phonon coupling strengths.

For a dilute Bose gas the mean square fluctuations in modal occupation probabilities are given as $\langle (\Delta n_i)^2 \rangle = \langle n_i \rangle + \langle n_i \rangle^2$.[57] This indicates that for low average occupation probabilities, $\langle n_i \rangle$, the fluctuations in mode $i$ are Poisson distributed, but for high $\langle n_i \rangle$ there exist positive correlations, *i.e.* "bunching".[58] To observe phonon bunching, a time record of a specific *modal* occupation probability is necessary, not simply a temporal average over (appropriately normalized) jump events. Monitoring the time correlations between events involving one specific jump size may enable such mode-specific temporal measurements, given the correspondence between the $(\Delta T)_i$ and $\omega_i$. Upon changing temperature from the regime where $\lambda_{dom} > 2\pi/q_i$ to that where $\lambda_{dom} < 2\pi/q_i$ ($q_i$ is the phonon wavevector), thermal depopulation of $\langle n_i \rangle$ occurs; this should be accompanied by a crossover from "bunched" to Poisson statistics. However, for the reasons noted earlier, interpretation of the $(\Delta T)_i$ when $\langle n_i \rangle$ are appreciable poses some complications.

## 10. Phonon Transport at the Quantum Limit: Energy Relaxation.

For macroscopic samples, we are accustomed to expect an exponential temperature decay after delivery of a small heat pulse. This classical behavior occurs when the system's thermal relaxation time is energy-independent, and holds in experiments: *a)* that are carried out on energy scales where the density of states is, in effect, continuous, and *b)* where there is essentially no energy dependence to the thermal pathway (conductance) enabling the system to equilibrate with its surroundings. In mesoscopic systems at low temperatures *both* should exhibit granularity. The weak residual coupling to the environment through "phonon waveguide" structures is highly energy dependent.[59] Under these conditions thermal relaxation temporally evolves in a non-exponential manner.[60] In fact, in the regime where cavity occupation numbers are small and the environmental coupling is very weak, *phonon shot noise* — reflecting phonon-by-phonon decay — may be observable.[61]

Perhaps the most intriguing of possible measurements are those involving temporal cross correlations between a pair of transducers coupled to the same nanocalorimeter. Above ~220mK the phonon exchange rate, $\tau_{ph}^{-1}$, is slow compared to both the phonon-electron and electron-phonon scattering rates, $\Gamma_{pe}$ and $\Gamma_{ep}$ (*cf.* traces 2 and 3 of Fig. 5). In this regime, as previously mentioned, the temperature fluctuations of the electron gases of the transducers should become anticorrelated. This reflects the fact that energy exchange between the pair of transducers, which is mediated by cavity phonons (the only energy pathway they share), occurs on a time scale faster than the phonon exchange with the environment.

## 11. Summary.

Detection of individual phonons by means of a nanoscale calorimeter appears feasible. When achieved, experiments analogous to those of quantum optics will emerge in mesoscopic phonon systems.[62] I have addressed concerns that might preclude attainment of this fascinating domain; it is up to experiments in progress to demonstrate whether these can be circumvented.[63] It is also clear that open theoretical questions, especially those regarding the quantum dynamics of energy exchange within nanoscale systems, must be addressed to interpret the proposed experiments. Irrespective of what ultimately proves experimentally possible, it seem clear that pursuing the quantum limit of thermal transport will lead to intriguing discoveries in nanoscale physics at ultralow energy scales.

**Acknowledgements.**

I thank Prof. John Worlock many enjoyable conversations that were crucial to the development of these ideas, and Dr. Keith Schwab for important input and critical comments on this manuscript. Many others also deserve thanks for their helpful insights, especially M.C. Cross, T.S. Tighe, M.B. Ketchen, B. Yurke, H.J. Kimble, A.N. Cleland, R. Lifshitz, J.P. Eisenstein, D.L. Goodstein, P. Mohanty, W. Wegscheider and Y.-C. Kao. I thank D.A. Harrington for carrying out the simulations of Fig. 8. Finally, I gratefully acknowledge support, in part, from the NSF through grant DMR-9705411.

[56] Their precise significance, however, can only be assessed through calculations that accurately model the experimental details, especially geometry.

[57] L.D. Landau and E.M. Lifshitz, ***Statistical Physics***, (Pergamon Press, Oxford, 1969) p. 355.

[58] See, *e.g.*, *"Bolometers for infrared and millimeter waves"*, P.L. Richards, Journ. Appl. Phys. **76**, 1 (1994).

[59] *"Acoustic phonon modes of rectangular quantum wires"*, N. Nishiguchi, Y. Ando, and M.N. Wybourne, J. Phys. Cond. Matter **9**, 5751 (1997). See also Ref. 2 and 3.

[60] See Ref. 2.

[61] For comparison, an experiment leading to the observation of the *photon-by-photon* decay within an optical cavity has been proposed and modelled. (*"Quantum nondemolition measurement of small photon numbers by Rydberg-atom phase-sensitive detection"* M. Brune, S. Haroche, V. Lefevre, J.M. Raimond, and N. Zagury, Phys. Rev. Lett. **65**, 976 (1990). Here what I propose is not non-destructive counting of cavity phonon number as in their proposal — but instead the observation of discrete jumps in the transducer temperature associated with cavity phonon annihilation. As discussed, when cavity occupation numbers are small this may be tantamount to "counting" arriving phonons. However interpretation of experiments will clearly require detailed theoretical modeling of the quantum dynamics.

[62] Experiments demonstrating "squeezing" of phonons have already been reported. See *"Vacuum squeezing of solids: Macroscopic quantum states driven by light pulses"*, G.A. Garrett, A.G. Rojo, A.K. Sood, J.F. Whitaker, and R. Merlin, Science **275**, 1638 (1997).

[63] Work of this nature is being carried out at Caltech by K. Schwab, C.-W. Fon, E. Henriksen, J.M. Worlock, and this author.